\documentclass[preprint,aps,12pt,amsmath]{revtex4}
\usepackage{graphicx}

\begin{document} 

\title{ZM theory I: Introduction and Lorentz covariance}

\author{Yaneer Bar-Yam}

\affiliation{New England Complex Systems Institute \\ 24 Mt. Auburn St., Cambridge, Massachusetts 02138}

\begin{abstract}
We consider defining time as a function of a cyclical field, an abstraction of a clock. The definition of time corresponds to a novel interpretation of the relationship between space-time coordinates of observers at different locations in space. As a first test of the utility of this definition, we show that it leads to a Lorentz covariant description of space-time. This derivation of Lorenz covariance provides a starting point for considering more general constructions that relate to physical laws. The definition of time couples time to space, making time not orthogonal to space, and making dynamics a result of geometry, providing a vehicle for curved space-time theories that generalize general relativity.
\end{abstract}


\maketitle

\section{Overview}   

In this paper we begin an investigation of the implications of a novel assumption about the relationship of fields, space and time. As in general relativity, the space-time metric varies from location to location, and the task is to describe this variation. According to the central assumption, time is not extrinsically defined either through a dynamical rule, or as an independent dimension of space-time. Instead, time can be obtained from the values of a kind of field variable. This treatment of the field variable $Z$ over the space manifold $M$ leads to ZM theory, which will be developed through a sequence of papers that explore 
the compatibility of the approach with conventional physical laws, principles, mechanics, elementary forces and excitations. This paper is largely devoted to introducing the basic concepts and showing that Lorentz covariance follows from the assumptions.  Other explorations of unconventional interpretations of time exist in the literature, e.g. the assignment of local times to groups of particles that follow quantum mechanics, with a separate treatment of general relativity in orthogonal coordinates.\cite{Kitada}

\section{Introduction}   

We consider a system, $Z$, with some set of distinctly labeled states that perform sequential transitions in a cyclic pattern, i.e. an abstract clock (similar to a conventional ``non-digital" clock consisting of a numbered dial, here with a single moving hand). Discreteness of the clock will not enter into the discussion in this paper. The clock states can therefore be extended to cyclical continuum, $U(1)$. The state change of the clock constitutes proper time, $\tau$, as defined by the clock. Since the clock is cyclical it is possible to represent the changing state using an oscillator language: 
\begin{equation}
\psi  = \exp ( - im \tau ), 
\end{equation}
where 
\begin{equation}
m = 2\pi /T
\end{equation}
is the cycle rate in radians. The clock phase is $c = m \tau$ modulo $T$, though this expression is not analytic so that derivatives should be defined in terms of $\psi$. 
The notation is chosen anticipating that $m$ will become the `rest mass' of the clock when it is reinterpreted as a particle. 
{\em A-priori} there is no difference between clockwise and counter-clockwise rotation, however, once one is identified it is distinct from the other.

As in traditional textbooks, we assume the clock time may be observed at any point in the environment. To introduce the space manifold, $M$, we consider a parameter, $x$, associated with the environment such that properties of the environment may lead to variation of the clock state with $x$ (we assume $x$ is a real number parameter). The distinctness of clock states is a kind of symmetry breaking, and the variation of clock state with space cannot be neglected as it would be if only topology mattered. The breaking of symmetry implies the reference value (coordinate origin) of clock time cannot be redefined arbitrarily. This treatment is different from the conventional description of space-time and is perhaps the most crucial defining assumption of ZM theory with other assumptions flowing more or less naturally from it. In the following we will be meticulous about providing details of derivations to clarify the assumptions.

As a first case to study, we consider the clock variation with $x$ to be uniform in $x$. This is not a fundamental assumption, but a convenient simple case, i.e. it will correspond to 
a single particle in a flat space (as traditionally understood). In general the approach is to develop formalisms that can describe progressively more elaborate variations. In this paper we consider only the case of uniform variation. We write the rate of change of the clock in space $x$ as $-k$, where the negative sign indicates that for positive $k$ proper time decreases for increasing $x$ values (see Figure 1). In order to properly define the rate of change of the clock in space, the units of distance measuring $x$ must be compared to the units of $\tau$. We assume, for convenience, that it is possible to choose the units measuring $x$ to be the same as the units measuring $\tau$, and thus the units of $k$ are the same as $m$.

\begin{figure}
\includegraphics[width=12cm]{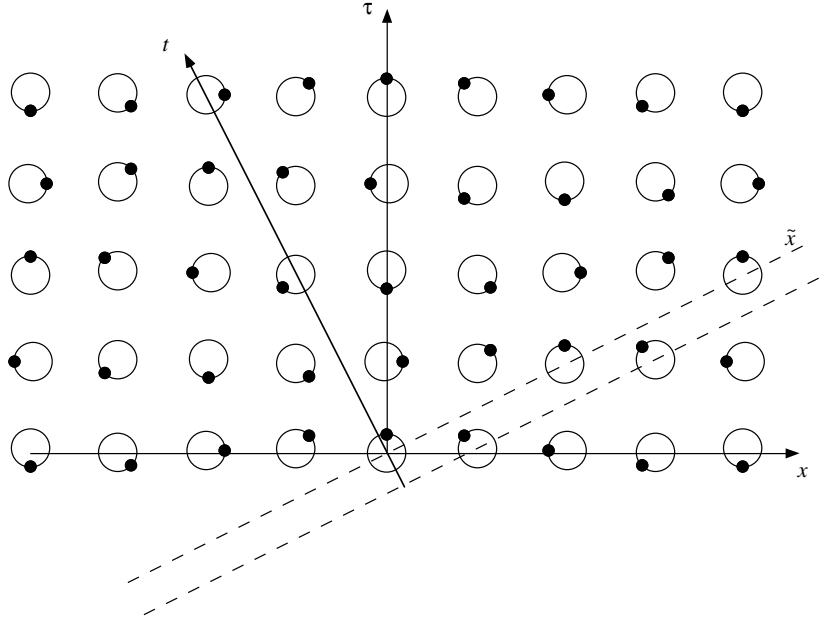}
\label{fig1}
\caption{Illustration of the clock state in
space-proper time dimensions, including a constant spatial variation
of proper time decreasing to the right. Note that the proper time
dimension is cyclical though it is illustrated here as unwrapped. Note also the rate of change of the clock phase along the time direction which is defined as $\omega$.
The diagonal dashed lines (marked by $\tilde x$) indicate lines of simultaneous clock time. }
\end{figure}

Our primary concern is the definition of time, $t$, as defined by the observer
which is not the same as the clock state. Significantly, we assume a specific relationship between observed time in relation to the intrinsic clock state change and clock state change in space.
The assumption will only be justified by the implications that are obtained in this and subsequent papers.  Specifically, observer time can be obtained by the rate of change of the clock, $c$, as a local gradient of the clock state in the two dimensional space given by $\hat \tau$ and $\hat x$ treated as a Euclidean space, whose direction can be considered as a `direction of time:'
\begin{equation}
(m, - k) =  m \hat \tau  - k\hat x.
\label{clock}
\end{equation}
We assume that time measures Euclidean distance along the time direction in the same units as $x$ and $\tau$, so the rate of change of the clock phase in the direction of time is 
\begin{equation}
\omega  = \sqrt {m^2  + k^2 }, 
\label{omega}
\end{equation}
i.e. consider possible directions of time  
\begin{equation}
\hat s =  \cos (\zeta )\hat \tau + \sin (\zeta )\hat x , 
\end{equation}
where $\zeta $ is the angle of rotation of $\hat \tau$ into $-\hat x$, with a clock change 
\begin{equation}
\begin{array}{ll}
d_s c &= \hat s \cdot (\partial _{\hat \tau}  c,\partial _{\hat x} c) \\
&= m \cos (\zeta ) + k\sin (\zeta ); 
\end{array}
\end{equation}
maximizing this with respect to the angle $\zeta $ gives 
\begin{equation}
\tan (\zeta ) = k/m, 
\label{geometry}
\end{equation}
and 
\begin{equation}
\omega  = d_t c = \sqrt {m^2  + k^2 }, 
\label{omegatime}
\end{equation}
where $\omega$ is defined to be the magnitude of the rate of change of the clock phase.

Distances are measured by considering variation in the $\hat \tau$ and $\hat x$ axes to be along independent Euclidean dimensions. Care must be taken to avoid confusion between the distance along the $\hat \tau$ axis and the value of the $\tau$ variable representing the clock rotation. By geometry, from Eq. (\ref{geometry}) the triangle in Fig. 2 has a vertical arm along the $\hat \tau$ axis of length $\tau_0$, 
a horizontal arm of length 
\begin{equation}
(k/m)\tau_0,
\label{x0}
\end{equation}
and a diagonal along the time direction of length 
\begin{equation}
\tau_0 \sqrt{1+k^2/m^2} = (\omega/m) \tau_0.
\end{equation}
Since time measures distance along the time direction this is the value of time at the point at the end of the diagonal:
\begin{equation}
t = (\omega/m) \tau_0.
\label{ttau00}
\end{equation}
We can specify locations in the two dimensional space by their coordinates along the axes 
\begin{equation}
(\tau_0,x_0)=\tau_0 \hat \tau + x_0 \hat x . 
\end{equation}
At the point along the time axis at the end of the diagonal we have
\begin{equation}
(\tau_0,x_0) 
=(\tau_0,-(k/m)\tau_0) 
\end{equation}
The value of the clock at this location can be directly calculated from the specification of the rate of its variation in the two dimensions in Eq. (\ref{clock}) as:
\begin{equation}
\begin{array}{ll}
c&=m\tau_0-k x_0 \\
&=m\tau_0 + (k^2/m) \tau_0 \\
&=(\omega^2/m) \tau_0 .
\end{array}
\end{equation}
$\tau$ is given by $c=m\tau$, so that 
\begin{equation}
\tau = (\omega^2/m^2) \tau_0
\label{tautau0}
\end{equation}
These expressions hold for $\tau_0$ and $\tau$ smaller than a cycle, or by analytic continuation for all real values, for the case of linear variation of $\tau$ in space.

\begin{figure}
\includegraphics[width=12cm]{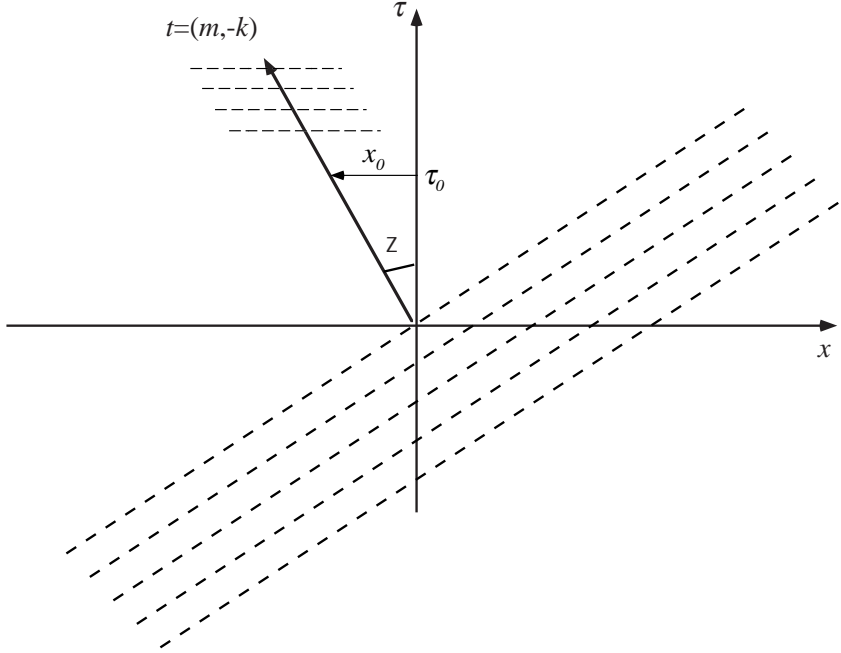} 
\label{fig2}
\caption{Illustration of the non-orthogonality of observer defined
time ($t$), and space ($x$). The angle $\zeta$ of the time axis is given by the
dependence of clock (proper) time $\tau$ on space $x$ as specified
by $k$. The horizontal dashed lines along the time axis
illustrate the lines of constant time parallel to the $x$ axis. The
diagonal dashed lines are at simultaneous clock time.}
\end{figure}

By the definition of the direction of time, and the rate of change of the clock along this direction, Eq. (\ref{omegatime}), we have that along the time direction:
\begin{equation}
\omega t = c = m \tau.
\label{timeaxis}
\end{equation}
This expression, however, does not apply along other directions. Indeed, the definition of time thus far only  identifies the variation of time starting from a specific initial location, e.g. $x=0$. If we want to relate time starting at different coordinates $x$, we must add an additional assumption. One possible and reasonable (though not necessary) assumption is that the observer time $t$ is synchronous along observer defined space coordinate $x$. Using this assumption, we can then consider time $t$ in the two dimensional space $(\tau_0,x_0)$. To express the dependence we can write the value of time as a function of proper clock time and position, where the variables $\tau$ and $x$ are not the same as the coordinate axes due to the variation of the clock along the $x$ axis. Treating $\tau$ as extended by analytic continuation, we can write the relationship: 
\begin{equation}
\omega t = m \tau  + kx. 
\label{trelation}
\end{equation}
The addition of the term $kx$ cancels the variation of the clock given by $c = m \tau$, which is  $-kx$ along the $x$ direction, so that time has the same value (0) along the $x$ axis, and a constant value along any parallel line. This implies that the value of time only varies perpendicular to the $x$ axis, i.e. along the $\hat \tau$ direction. The coefficient of time, $\omega$, on the left side of Eq. (\ref{trelation}) is chosen for consistency with the variation of time along the time axis. Combining Eq. (\ref{timeaxis}) with Eq. (\ref{trelation}) we see that $x=0$ along the time axis, consistent with the intuitive meaning of a direction of time as the axis of the origin of spatial coordinates. 

Thus, the direction of time, along with the assumption of defining time so that space is synchronous, implies the observer's origin for the coordinate $x$ shifts to the left by a displacement that depends on time and proper time. The length of this displacement was obtained directly from the geometry of the triangle in Fig. 2 in Eq. (\ref{x0}). To write this displacement in terms of time we relate time to $\tau_0$. Eq. (\ref{ttau00}) gives the relationship of time and $\tau_0$ along the time direction as 
\begin{equation}
t = (\omega/m) \tau_0.
\label{ttau0}
\end{equation}
Since both $t$ and $\tau_0$ do not change along the $x$ axis, this equation is valid everywhere in the plane. Thus the horizontal displacement of the axis origin (and any other mark on the axis) is given by 
\begin{equation}
(k/m)\tau_0 = (k/\omega) t .
\label{vrelation}
\end{equation}
Since the coordinate axis is translating to the left, we might also say that the observer perceives a reference location to translate to the right with a velocity \begin{equation}
v = k/\omega,
\end{equation}
so that we can write
\begin{equation}
\begin{array}{ll}
x&=x_0+(k/m)\tau_0 \\
&=x_0+(k/\omega) t.
\end{array}
\end{equation}
This identifies a connection between the velocity in special relativity and $k$ in ZM theory. 

There are a number of comments that follow from the discussion of the observer clock relationship in ZM theory. These comments help in explaining the relationship of these ideas to special and general relativity and may enable further development. Still, these interpretive comments are not essential to the mathematical formalism that has been developed above. 

First, it is important to recognize that the ability to observe a clock throughout space does not mean that time as defined by different observers is the same, or the same as time defined by the clock. Indeed, the focus of ZM theory, similar to special and general relativity, is on relating different observers' concepts of space-time. 

Second, the approach of ZM theory in considering the relationships between observers and their clocks is similar to special relativity in which the clock rates of two observers moving in relation to each other are not the same, so that one observer can observe that the clock of the other observer does not report the same time as his/her own clock, i.e. as the clock that is stationary in his/her reference frame. 

Third, unlike special relativity, where clocks vary in velocity and arbitrarily overlap in space, in ZM theory time is identified with the values of the field variables in a particular region of space.  The association of the behavior of time with a region of space is analogous to the approach of general relativity, in which the metric of space-time is associated with each location of space, and curvature of space-time is related to the local stress-energy-momentum tensor.\cite{LL} In ZM theory time is more directly defined through its dependence on the  field variable $Z$, i.e. the clock. 

Fourth, more specifically, consider two regions of space, $A$ and $B$. In ZM theory, an observer at any specific location of observation reports that the clock state is $\psi  = \exp ( - im \tau )$. Our concern, however, is how the observer in region $A$ reports the behavior of the clock in region $B$. In order to report on the behavior of the clock, the observer in region $A$ may extrapolate his own local definition of time and space, which we might denote $(t_A,x_A)$, but which we write more simply as $(t,x)$, to the region of space $B$. Since this is an extrapolation, the behavior of space-time in region $B$, as defined by the observer in region $B$, may not be the same as that defined by the observer in region $A$. We therefore consider various possible clock behaviors in region $B$ and how they are to be understood by the observer in region $A$.  In particular, a key way that the extrapolated space-time may be different from that defined by the observer at $B$ is that the clock at $B$ may not be synchronous across the extrapolated space $x$ at $B$. It is the extrapolated time $t$ which is synchronous across the extrapolated space $x$. Thus, the formal development we have given for space-time $(x,t)$ as defined by an observer can be understood directly as the space-time that is extrapolated from region $A$ to region $B$. 

Fifth, the non-orthogonality of space and the direction of time as seen by the observer gives rise to mathematical and conceptual problems. ZM theory associates their resolution to physical laws, interactions and dynamics. Issues include the difference between observer defined time and a particular clock's time as measured by the observer, and spatial variation in the clock/observer time. 

Sixth, we emphasize the dissonance between various considerations of time. Specifically, the rate of change of the clock at a location $x$ does not correspond to the rate of change of time ($d_t c\ne 1$), and yet the rate of change of time is not independent of the rate of change of the clock, being defined in terms of (or related to) the rate of change of the clock in the combined space and proper time. 

Seventh, the cyclical nature of $c$ and its non-analyticity across cycle boundaries are key features of ZM theory. 
The equations above are valid everywhere locally in a small enough neighborhood, and for an appropriate choice of the location of the cycle discontinuity. For broader applicability we should write  
\begin{equation}
 i d_t \psi  = \omega \psi .
\end{equation}  
From the perspective of the observer, who considers space and time to be independent variables, the clock performs a space-time rotation that can be represented as: 
\begin{equation}
\psi = \exp (i(kx - \omega t)). 
\end{equation}  
At any location this gives the same values as  
\begin{equation}
\psi = \exp(-im\tau ). 
\end{equation}  
In later papers of this series these issues will play a role in quantum phenomena. Assuming we are describing variation within a single cycle, or that analytic continuations are valid, enables the description of classical theoretical frameworks. Because the definition of time is in terms of the clock, the issue of analyticity arises even in the definition of coordinate systems. In this paper, because we are considering only linear variations of the clock in space, such analytic continuations are valid throughout space. We therefore discuss coordinates assuming the cycling of $\tau$ can be ignored. The conditions for this assumption will be discussed in a later paper.

\section{Lorentz covariance}

We consider the set of possible representations and how they can be transformed to each other. Since a representation is characterized only by the spatial variation of the clock, we can label two representations by $k$ and $k'$. Alternatively, using the fourth comment in Section II (introducing the association of observers with regions of space), we can consider two observers in regions $A$ and $A'$ who have different views of the clock at $B$ and therefore report different values of spatial variation of the clock $k$ and $k'$. 

Lorentz covariance is the ability to relate valid descriptions by different observers using Lorentz transformations. To demonstrate Lorentz covariance, we ask what redefinition of coordinates $(t,x)$ will give rise to a change from one clock representation with spatial variation $k$, to one with $k'$. A representation is given by the variation of $\tau$ in space-time, so we solve Eq. (\ref{trelation}) for $\tau$ to obtain:
\begin{equation}
\tau(t,x)=(\omega t - k x)/m. 
\label{tau}
\end{equation}
In order for Lorentz covariance to apply, the same equation must hold after redefinition of coordinates with the spatial variation $k'$, i.e. for a second observer, with different definitions of space-time:
\begin{equation}
\tau(t',x')=(\omega' t' - k' x')/m. 
\label{tauprime}
\end{equation}
It is necessary to show that the Lorentz transformation is consistent with this relationship. 
Writing the new coordinates as linear combinations of the old: 
\begin{equation}
\begin{array}{ll}
x' &= \gamma_r (x + v_r t), \\
t'& =  \gamma_r (t + v_r x),
\end{array}
\end{equation}
Inserting the coordinate transformation into Eq. (\ref{tauprime})  and equating to Eq. (\ref{tau}) we have:
\begin{equation}
\omega t - k x = \omega'  \gamma_r (t + v_r x) - k'  \gamma_r (x + v_r t)
\label{consistency}
\end{equation}
which must be valid for all $(x,t)$.  For completeness we include in detail the solution of this
equation, which we obtain by equating the coefficients of $x$ and $t$, giving: 
\begin{equation}
\begin{array}{ll}
\omega &= \omega' \gamma_r  - k' \gamma_r v_r, \\
-k & = \omega' \gamma_r v_r- k'  \gamma_r ,
\label{keyequations}
\end{array}
\end{equation}
Dividing the second equation by the first,
\begin{equation}
 - k / \omega =  (\omega' v_r- k' ) /  (\omega' - k'  v_r)
\end{equation}
defining $v=k/\omega$ and $v'=k'/\omega'$, we have
\begin{equation}
 - v =  (v_r - v' ) /  (1-v_r v' ).
 \end{equation}
Solving for $v_r$ gives
\begin{equation}
 v_r =  (v' - v ) /  (1-v v' ).
 \label{velocityaddition}
 \end{equation}
Which is the usual Lorentz velocity composition formula for velocities in a single dimension. To solve for $\gamma_r$ we rewrite Eq. (\ref{keyequations}):
\begin{equation}
\begin{array}{ll}
\omega / k'  &= \omega' \gamma_r / k'  - \gamma_r v_r, \\
-k / \omega'& = \gamma_r v_r- k'  \gamma_r / \omega'.
\end{array}
 \end{equation}
Adding the equations yields
\begin{equation}
\omega / k'  -k / \omega' =  \omega' \gamma_r / k' - k'  \gamma_r / \omega'.
 \end{equation}
Solving for $\gamma_r$ gives
\begin{equation}
 \gamma_r = (\omega  \omega'  - k k' )/  (\omega'^2 - k'^2 ).
 \end{equation}
 Using Eq.(\ref{omega}) in the primed coordinate system to simplify the denominator we have
 \begin{equation}
 \gamma_r = (\omega  \omega'  - k k' )/ m^2.
 \end{equation}
 Finally, defining  $\gamma = \omega / m$ and $\gamma' = \omega' / m$ we obtain:
 \begin{equation}
 \gamma_r =\gamma  \gamma'  (1- v v' ).
 \label{gammaaddition}
 \end{equation}
This expression can either be recognized or shown by additional algebra to be equivalent to the usual Lorentz transformation expression for $\gamma$ composition.\cite{Jackson} Thus we have demonstrated Lorentz covariance, with the identification of $v_r$ and $\gamma_r$  according to Eqs. (\ref{velocityaddition}) and (\ref{gammaaddition}), which are the usual Lorentz expressions for velocity and $\gamma$ composition. 
 
It is possible to simplify significantly the above derivation of Lorentz covariance. The well known composition rules of the Lorentz transformation imply that it is sufficient to consider the case where the primed coordinate system has $k' = 0$; then the composition properties imply that we can more generally transform between $(x,t)$ and $(x',t')$ with arbitrary values of $k'$. In the simpler case, setting $k' = 0$, we have $\omega' = m$ and:
\begin{equation}
\tau(t',x') =t'.
\end{equation}
In this case, 
the consistency equation (Eq. (\ref{consistency})) becomes
\begin{equation}
\omega t - k x = m \gamma_r (t + v_r x) .
\end{equation}
Equating the coefficients of $x$ and $t$, gives:
\begin{equation}
\begin{array}{ll}
\gamma_r &= \omega / m \\
v_r &= - k / \omega,
\end{array}
\end{equation}
the standard expressions for the Lorentz transformation. The negative sign arises because we are transforming from the representation $k$ to $k'=0$. Thus, Lorentz covariance has been demonstrated for the assumptions of ZM theory.

There are a number of comments that follow from the derivation of Lorentz covariance. These interpretive comments are not essential to the mathematical formalism that has been developed above. 

First, just as in special relativity, no mechanism was given to achieve a Lorentz transformation as this would require acceleration. Acceleration is not included at this level of description.

Second, comparing with the conventional treatment of special relativity we see that the starting point for ZM theory is different: (a) the concept of equivalent inertial frames does not appear as a fundamental principle, (b) the concept of light and its invariant velocity does not appear as a fundamental principle, (c) the ideas of signal propagation and causality are not discussed, (d) the invariance of the space-time interval does not explicitly appear, (e) the extrinsic variable in special relativity is the velocity, $v$, whereas in ZM theory it is $k$. Perhaps more significantly, the geometry of the transformation between the two coordinate frames appears different, not the least because of the non-orthogonality of $x$ and $t$. Still, the appearance of the Lorentz transformation means that according to some perspective special relativity and ZM theory are equivalent, or have a kind of correspondence, for flat space. Thus, Lorentz covariance suggests many of the properties of special relativity may also apply to ZM theory, but they do not appear in the assumptions.


Third, despite the mathematical mixing of $\tau$ and $x$, they play different roles from each other in this formalism and, together, they are different from the conventional abstract notion of space-time frames of observation in special and general relativity.  $\tau$ is a cyclical field coordinate, while space is introduced in a more conventional way as an extrinsic real valued coordinate, though one whose definition may vary from location to location as in general relativity. Nevertheless, Lorentz covariance follows from the assumptions as given.  

Fourth, the use of a transformation that mixes $x$ and $\tau$ raises the question as to whether there is a way that they can be treated on equal footing. Since space may also reflect a change of state, similar to $\tau$, despite the distinction in treatment between position and time variables it is possible that we could consider the Lorentz transformation to be a reapportionment of states between field and environment, i.e. between $\tau$ and $x$, in a way that would place them on equal footing. While this approach may be fruitful, it will not be pursued here.

Fifth, the variability of the clock over space 
implies that space must be reinterpreted if it is defined by simultaneity of this clock. Choosing synchronous space to be defined by the clock state at every location leads to intervals in space including intervals in proper time. We define $\tilde x$ (see Figure 1) as the space for the synchronous clock. The direction of the axis is given by
\begin{equation}
(k, m) =  k \hat \tau  + m \hat x.
\end{equation}
The values of $\tilde x$ can be written in terms of $x$ and $\tau$. Assuming that $\tilde x$ is the same on the $\tau$ axis, and the units of length are unchanged, implies 
\begin{equation}
\omega \tilde x = mx - k\tau.
\end{equation}
As long as we are considering how the observer (at $A$) describes the behavior of the clock over space (at $B$) we will continue to use the space $x$, rather than $\tilde x$.  $\tilde x$ will enter when we consider explicitly the variable metric of space in general relativistic formalizations in a future paper.

Sixth, in ZM theory time should be interpreted, as we experience it, as a sequence of state changes (whether discrete or continuous does not matter at this point) that we can identify as associated with one or more clocks. Time is not, {\em a-priori}, an extended dimension. This is consistent with our inability to observe events along time in the same way as we observe events at multiple locations of space. However, it may be possible under some circumstances to associate a dimension with the sequentiality of state changes, as we have done above in the definition of time. Such extensions of time may not, however, be possible in general.

Seventh, in conventional quantum mechanics and quantum field theory,\cite{Huang} space and time are extrinsic parameters. Still, in the Dirac formulation of quantum mechanics\cite{Dirac} the state of the system $\left| {} \right\rangle $ is a distinct entity from the spatial wave function defined as $\left\langle {x} \left |   \right.  \right\rangle $. While eventually in that formalism $x$ is considered as associated with the particle coordinate rather than as the definition of space, still, as a conceptual framework, we might consider the approach taken here to follow this prescription by separately identifying $\left\langle x \right|$ as a property of the observer/environment rather than of the system, i.e. space is a characteristic of the environment. The observed system has a globally (i.e. universally in space) set of degrees of freedom to which the environment may make contact. This is {\em a-priori} consistent with the non-locality of quantum states and transitions. Causal propagation is not assumed but is presumed to arise in an appropriate limit. Time, which starts as a property of the system rather than of the environment---through the transitions of the clock (proper time), can become mixed with space. The geometric mixing of time and space (distinct quantities describing environment and system relationships) is then perceived as dynamics.

\section{Conclusions}

In this paper we have discussed how an unusual definition of time can result in Lorentz covariant description of flat space-times, as considered in special relativity. In subsequent papers of this series we will develop this formalism to show how defining time as a local function of field variables can give rise to a variation of properties in and of space yielding physical laws, including dynamics.

\end{document}